\documentclass[sigconf]{acmart}
\AtBeginDocument{%
  \providecommand\BibTeX{{%
    \normalfont B\kern-0.5em{\scshape i\kern-0.25em b}\kern-0.8em\TeX}}}

\usepackage{breakcites}
\usepackage{microtype}
\usepackage{graphicx}
\usepackage{subfigure}
\usepackage{booktabs} % for professional tables

\usepackage{amssymb,amsmath}
\usepackage{pifont}
\usepackage{color}
\usepackage[english]{babel}
\usepackage{array}
\usepackage{bm}
\usepackage{multirow}
\usepackage{enumitem}
\usepackage{pifont}
\usepackage{tikz}
\usepackage{amssymb}
\usepackage{enumitem}
\usepackage[english]{babel}
\usepackage{array}
\usepackage{bm}
\usepackage{multirow}
\usepackage{enumitem}
\usepackage{pifont}
\usepackage{graphicx}
\newcommand{\eat}[1]{}
\usepackage{flushend}
\usepackage{makecell}
\usepackage{pifont}
\begin{document}

\title[QuGeo: Quantum Learning for Geoscience]{QuGeo: An End-to-end Quantum Learning Framework for Geoscience --- A Case Study on Full-Waveform Inversion}

\author{Weiwen Jiang}
\affiliation{%
  \institution{Electrical and Computer Engineering Department, George Mason University}  
}
\author{Youzuo Lin}
\affiliation{%
  \institution{Los Alamos National Laboratory}
  \institution{University of North Carolina at Chapel Hill}  
}

\renewcommand{\shortauthors}{W. Jiang and Y. Lin}

\begin{abstract}
The rapid advancement of quantum computing has generated considerable anticipation for its transformative potential. However, harnessing its full potential relies on identifying ``killer applications''. In this regard, \textit{QuGeo} emerges as a groundbreaking quantum learning framework, poised to become a key application in geoscience, particularly for Full-Waveform Inversion (FWI). This framework integrates variational quantum circuits with geoscience, representing a novel fusion of quantum computing and geophysical analysis. This synergy unlocks quantum computing's potential within geoscience. It addresses the critical need for physics-guided data scaling, ensuring high-performance geoscientific analyses aligned with core physical principles. Furthermore, QuGeo's introduction of a quantum circuit custom-designed for FWI highlights the critical importance of application-specific circuit design for quantum computing. 
In the OpenFWI's FlatVelA dataset experiments, the variational quantum circuit from QuGeo, with only 576 parameters, achieved significant improvement in performance. It reached a Structural Similarity Image Metric (SSIM) score of 0.905 between the ground truth and the output velocity map. This is a notable enhancement from the baseline design's SSIM score of 0.800, which was achieved without the incorporation of physics knowledge.
\end{abstract}

\maketitle
\vspace{-5pt}
\section{Introduction}
\vspace{10pt}
\noindent Quantum computing's integration with machine learning is a rapidly expanding field, though it is still in search of groundbreaking applications to fully showcase its capabilities. In geophysical inversion, critical for areas like civil infrastructure and energy exploration, recent advancements have applied quantum annealing to seismic inversion problems. These efforts, however, primarily tackle linearized versions of the problem. For instance, Souza's approach reformulates seismic inversion into linear equations and QUBO formulations~\cite{Souza-2022-Quantum}, and Greer develops a QUBO-compatible linearized inversion for dual velocity values, both based on traditional seismic techniques~\cite{Greer-2023-Early}. In contrast, our work introduces the first learning-based seismic inversion technique on general-purpose quantum computing platforms, marking a novel direction in this domain.

Characterizing subsurface geology is essential for a range of applications, from earthquake research and civil infrastructure to new energy exploration and environmental studies. Seismic inversion, which reconstructs subsurface images from seismic waves, uses full-waveform inversion (FWI) to consider complete waveform data, including amplitude and phase. FWI, addressing non-linear challenges, offers enhanced accuracy and resolution compared to linear methods. Currently, FWI is approached through physics-based and machine learning-based methods. While physics-based methods are precise, they face computational challenges and issues like ill-posedness and cycle skipping. Conversely, machine learning-based methods, designed to address these issues, are promising but heavily dependent on extensive training data, which can result in high computational demands. 
What's more, seismic waves continuously received by sensors from different locations have a high correlation on both spatial and temporal.
Our study introduces quantum computing as a novel solution to reduce the computational burden in data-driven FWI and leverage fundamental quantum mechanics, particularly the entanglement, to extract highly correlated features, showcasing its potential to enhance both the efficiency and effectiveness of machine learning in seismic inversion.
% \ylnote{(Add another benefit here.)}

% Designing quantum learning for FWI presents unique challenges. 

Unique characteristics and structures of geophysical problems, however, set challenges in designing quantum learning for FWI. 
Firstly, there is a scarcity of standard datasets tailored to current quantum capabilities; careless design could result in data misaligned with physical realities, affecting performance. Secondly, a dedicated quantum design framework is lacking; straightforwardly using the quantum learning algorithm (a.k.a., Variational Quantum Circuit or VQC) can easily be suboptimal or inefficient because it cannot exploit the specific properties inherent in geophysics to optimize the design. 
Finally, the most effective ways to harness the full potential of quantum computing in this context remain unclear, requiring further exploration and innovation in the field.

In our work, we introduce QuGeo, an innovative quantum learning framework tailored to address seismic FWI challenges. To tackle the mentioned hurdles, firstly, we have created a physics-informed dataset via a governing wave equation, on top of which we further developed a classical machine learning-based data converter to perform data scaling according to the quantum resource constraints efficiently.
% for model training. 
Secondly, we developed an application-specific VQC, namely QuGeoVQC, to incorporate domain-specific knowledge in optimizing the design.
QuGeoVQC explores the designs of a data encoder and VQC computing structure to extract spatial and temporal features; it also exploits characteristics of FWI problem to simplify and optimize the encoder to boost the performance.
% We have also specifically designed QuGeoVQC, an application-specific quantum circuit, to enhance performance for seismic FWI. 
Furthermore, we present a novel data batching technique adapted for quantum computing, which enables quantum computers to process $N$ batches of data in parallel with only $logN$ additional qubits,
% with no computational overhead and 
further catering to the high computational demands from learning-based FWI and advancing the potential of QuGeo in seismic inversion.

The main contributions of this paper are as follows:

\vspace{-\topsep}
\setlength{\parskip}{5pt}
\setlength{\itemsep}{0pt plus 1pt}
\begin{itemize}
    \item To the best of our knowledge, this is the very first pilot work in exploring the capacity of general-purpose quantum computing for seismic inversion problems, showcasing the potential of quantum computing for geoscientific analyses.
    \item We unveil the essential need for a physics-informed dataset to ensure data is aligned with physical realities while simultaneously satisfying the resource constraint in current quantum computing platforms.
    \item We develop an application-specific variational quantum circuit to leverage domain-specific knowledge in optimizing the design, which is equipped with a novel data batch engine to underpin the high computational demand to process extensive data for learning-based seismic inversion.    
\end{itemize}
\vspace{-\topsep}

To evaluate the effectiveness of our proposed QuGeo framework, we implemented the proposed quantum learning innovations on TorchQuantum platform \cite{wang2022quantumnas}
and 
conducted tests using the widely recognized OpenFWI dataset from the geophysics community~\cite{deng2022openfwi}. 
These tests aimed to validate the integration of physics knowledge and assess the performance of our custom-designed QuGeoVQC module. Our findings demonstrate that the incorporation of physics knowledge in QuGeo not only leads to high prediction accuracy, as reflected in the SSIM values, but also results in a remarkably efficient learning model utilizing just 576 parameters.

This paper is structured as follows: Section 2 presents the background.
In Section 3,  we detail the QuGeo Framework.
Evaluation results and concluding remarks are presented in Sections 4-5.

\section{Background and Related Work}

\noindent\textbf{Data-driven Quantum Learning.} 
The core component of a typical quantum learning framework is a parameterized quantum circuit (a.k.a., variational quantum circuit or ansatz circuit).
Like classical machine learning, there are two steps in quantum learning: forward propagation and backward propagation.
The forward propagation only involves quantum computing, where data will be encoded, processed, and measured in the quantum circuit.
The backward propagation further involves classical computing to calculate the loss according to the measurement results and label, which will then be used to update the parameters in VQC.
The optimization framework is to find the optimal parameters in quantum circuits so that the cost function of a certain task can be minimized. 

\begin{figure}[t]
    \centering
    \includegraphics[width=0.48\textwidth]{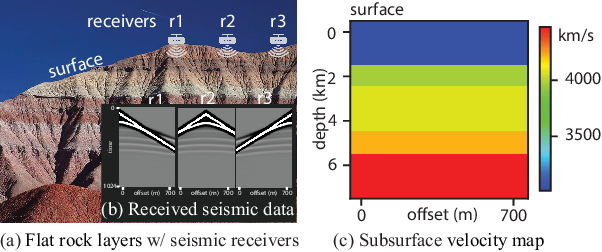}
    \caption{Illustration of the target geophysics application: (a) the photo of a flat rock layer; (b) the seismic waveform data obtained from the receivers placed on the surface; and (c) the velocity map used to characterize the subsurface structure.}
    \label{fig:mot1}
\end{figure}

Quantum learning has a wide of applications, such as Quantum Approximate Optimization Algorithm (QAOA) for combinatorial optimization problems \cite{wurtz2021maxcut}, Variational Quantum Eigensolver (VQE) for quantum chemistry and quantum simulations \cite{mihalikova2022cost}, and Quantum Neural Networks (QNN) for machine learning tasks \cite{dhotre2022exploring,wang2022quantumnas,jiang2021co,hu2023toward,hu2022quantum,wang2023qumos}.
% Recently, several QL algorithms have theoretically shown the potential quantum speedups over their classical counterparts on computation complexity in different algorithms: support vector machine \cite{rebentrost2014quantum}, quantum generative model \cite{gao2018quantum}, quantum kernel estimation \cite{liu2021rigorous}, and quantum binary neural network \cite{jiang2021co}.
% Most recently, \cite{huang2022quantum} experimentally shows the quantum advantage in learning patterns on a Google Sycamore processor.
% To our knowledge, this is the first study to explore quantum learning in the Geoscience FWI. 
Compared with classical computing, quantum learning gets benefits from entanglement, which has been proven to be scalable to learn from the exponentially increasing size of data \cite{sharma2022reformulation}.

\noindent\textbf{Geoscience application --- Full-Waveform Inversion (FWI).}
\textit{Subsurface structures} are typically layer-structured; an example is the flat rock layers, as illustrated in Figure \ref{fig:mot1}(a).
Understanding subsurface structures is a multidisciplinary effort with broad implications. 
In the field of new and renewable energy exploration, accurately characterizing subsurface structures is crucial for effectively locating energy sources below the Earth's surface.
Note that it not only contributes to resource exploration and extraction but also plays a vital role in addressing environmental challenges, ensuring infrastructure stability, and advancing our knowledge of Earth's geological processes.
\textit{Seismic data} and \textit{velocity maps} are two key elements to describe subsurface structures, discussed below.

\textit{Seismic data}, or seismic waveforms, are composed of waves that travel through the Earth's subsurface. These waves are often generated by controlled explosions or by striking the ground with heavy weights. Captured by geophones or receivers, these waves reflect varying subsurface layers. The collected data is then processed to create detailed waveforms in a shot gather, as illustrated in Figure~\ref{fig:mot1}b. From this shot-gather waveform data, a \textit{velocity map} can be derived, indicating the speed at which seismic waves travel through different subsurface media, as shown in Figure~\ref{fig:mot1}c. Since rock properties affect wave speed, the velocity map is essential in identifying and characterizing subsurface geological structures, such as rock layers, faults, and reservoirs. For instance, the example in Figure~\ref{fig:mot1}c displays a subsurface structure with five distinct layers, each representing different rock properties and depths.

Building on those concepts, we introduce \textit{Full-Waveform Inversion (FWI)}, a sophisticated seismic imaging technique widely used in geophysics for creating detailed and accurate subsurface models. FWI aims to minimize the discrepancy between observed and simulated seismic waveforms.  Applied extensively in both academic and engineering~\cite{Virieux-2009-Overview}, FWI essentially predicts the subsurface velocity map from seismic waveforms recorded at surface receivers. Traditionally, FWI relies on physics-based simulations that are computationally intensive. 
The concept of data-driven FWI, introduced in recent years \cite{wu-2019-inversionnet}, has led to its growing success.
In this work, we pilot a data-driven quantum learning for Geoscience FWI, harnessing the power of quantum computing to enhance the FWI.

\section{Q\MakeLowercase{u}G\MakeLowercase{eo} Framework}

\begin{figure}[t]
    \centering
    \includegraphics[width=0.4\textwidth]{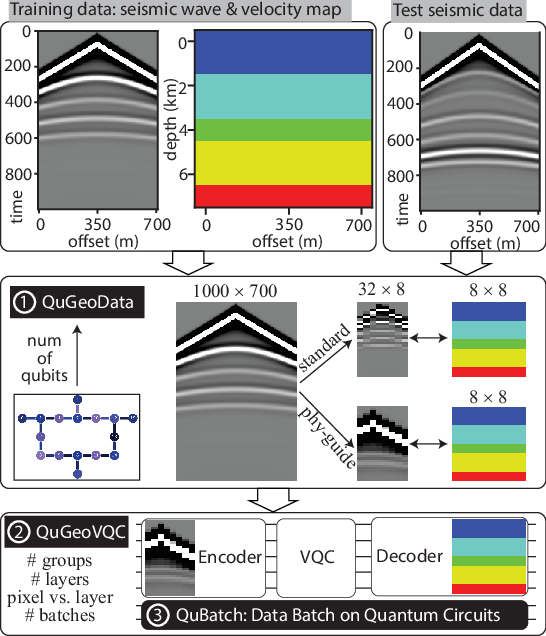}
    \caption{Illustration of the proposed \textit{QuGeo} framework.}
    \label{fig:framework}
\end{figure}

\noindent Figure \ref{fig:framework} shows the proposed \textit{QuGeo} framework, which is composed of three components: {\Large{\ding{172}}} \textbf{QuGeoData} will consider the quantum device's capacity and scale the seismic wave data appropriately; {\Large{\ding{173}}} \textbf{QuGeoVQC} is a computation engine which is key to achieving practical usage of near-term noisy quantum computers;
% which is designed to process the seismic data and generate the velocity map; 
and {\Large{\ding{174}}} \textbf{QuBatch} is a performance booster that provides the ability to process data in parallel to unleash the power of quantum computing.

\subsection{QuGeoData: Physics-Guided Data Scaling}

\subsubsection{Scaling data with velocity map.}\label{sec:q-d-fw}~\hspace{1pt}
We have pairs of seismic wave and velocity maps in the training dataset.
Instead of directly scaling seismic data, we propose to employ Forward Modeling to generate the seismic wave data from downsampled velocity maps.
The governing equation of the acoustic wave propagation in a 2-dimension isotropic medium with a constant density is as follows:
\begin{equation}
\nabla^2p-\frac{1}{c^2}\frac{\partial ^2p}{\partial t ^2}=s
\vspace{-8pt}
\label{Acoustic}
\end{equation}
where $\nabla^2=\frac{\partial ^2}{\partial x ^2}+\frac{\partial ^2}{\partial y ^2}$, $c$ is velocity map, $p$ is pressure field and $s$ is source term. Velocity map $c$ depends on the spatial location $(x,y)$ while the pressure field $p$ and the source term $s$ depend on the spatial location and time $(x,y)$ and $t$.
We will follow the Forward Modeling algorithm \cite{forwardmodeling}.
The seismic data is simulated using finite difference methods 
with the absorbing boundary condition.

\subsubsection{Scaling data without velocity map.}~\hspace{1pt} \label{sec:q-d-cnn} When the \textit{QuGeo} is applied to real-world applications, we do not have the subsurface velocity map, which requires the scaling of data without using the velocity map. 
In addition, it has the requirement of high efficiency in performing data scaling.
To address these issues, we propose an ML-based data scaling approach to train a convolution neural network (CNN) for data compression.
Our hypnosis is that physics-related features can be learned from the feature extraction by CNN.
% in the data compressed by the physics-guided Forward Modeling approch can be 

The key step is to build the dataset. Here, we have the original seismic wave data ($D$), which has the dimension of $1000\times 700$ as in the example from Figure \ref{fig:framework}.
We also have the physics-guided scaled data ($phyD$), which has the dimension of $32\time 8$ in the example.
The pair of data $\langle D, phyD\rangle$ will be built as the training dataset.
On top of the dataset, we designed a LeNet-like CNN to perform data compression, which contains two convolutional layers (including a ReLU function after the convolution operation) and a fully connected (FC) layer.
Although simple, it is quite effective to perform the data compression, as will be shown in the experimental results.

\subsection{QuGeoVQC: Quantum Circuit Design}

\subsubsection{Encoder of Multi-source Seismic Data.}~\hspace{1pt}
The encoding of seismic wave data to qubits is based on a state-of-the-art quantum encoder from \cite{li2023novel}, called spatial-temporal encoder ``ST-Encoder''.
It was originally designed for natural images, encoding a group of data (i.e., spatial-close $N$ data) to amplitudes of $log_2{N}$ qubits.

To use ST-Encoder in \textit{QuGeoVQC}, we need to identify how to map data to the amplitudes of a set of qubits. 
Seismic data has three dimensions, including (1) sources, which generate waves from the surface at different locations; (2) receivers, which receive waves on the surface; and (3) elapsed time, which shows the wave pressure received over time.
As shown in Eq. \ref{Acoustic}, one source represents an independent event (say vibration from vibroseis trucks), and each of them can be calculated by the differential of pressure on time and location.
Therefore, to better extract features from different sources, we group and encode data from the same source into qubits.

\subsubsection{Variational Quantum Circuit.}~\hspace{1pt}
We will follow ST-VQC \cite{li2023novel} to create sub-VQCs to independently process data in each group, and gradually commute between groups by using multi-qubit gates across different sub-VQCs.
There are a couple of design hyperparameters that need to be determined. 
% Specifically, 
% \textbf{\textit{Design space:}} 
In each sub-VQC, it is necessary to determine how many layers of VQC are needed, which reflects the number of trainable parameters.
% We need also to decide which gates are used in one layer.
In addition, the order of entanglement between different groups needs to be identified.

\subsubsection{Decoder of Velocity Maps.}~\hspace{1pt}
% Unlike the data encoder and VQC, the decoder has not been well optimized in the existing works.
In carrying out the waveform inversion task, we observe the simplification of the decoding circuit can significantly improve performance, which takes benefits of the task's property.
The decoder design is related to the definition of the loss function used in training VQC.
Let $G$ be the ground-truth velocity map, and $G_{i,j}$ the velocity at location $\langle i,j\rangle$, where $i,j\in [0,N)$ represents the coordinate.
A straightforward design is to conduct a pixel-wise comparison between ground truth and the forward results.
In this case, the decoder needs $N^{2}$ velocities from the quantum circuit, denoted as $D$.
For example, the mean squared error between ground truth and VQC outputs can be defined as:
\begin{equation}\label{eq:px}
loss_{pixel}=\sum_{i\in [0,N)}\sum_{j\in[0,N)}\{(G_{i,j}-D_{i,j})^2\}
\vspace{-10pt}
\end{equation}

As most subsurface has a flat structure, we can simplify the regression output data.
Specifically, instead of pixel-wise comparison, we can predict one velocity of each row. As a result, we only predict $N$ velocities, denoted as $D^{\prime}$, leading to the following MSE:
\begin{equation}\label{eq:ly}
loss_{layer}=\sum_{i\in [0,N)}\sum_{j\in[0,N)}\{(G_{i,j}-D^{\prime}_{j})^2\}
\vspace{-10pt}
\end{equation}
% In this way, the decoder only extracts $N$ data, reduced from $N^{2}$.

The above method can be generalized for the non-flat subsurface, such as curve structures.
Because the subsurface mediums between curves have the same material, indicating a similar velocity.
In this case, we can still use one velocity for each row but need to predict a multi-variable function to describe the curve.
The row velocity will be used for all locations between two curves.

\begin{figure}[t]
    \centering
    \includegraphics[width=0.48\textwidth]{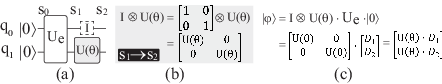}
    \caption{Illustration of the proposed \textit{QuBatch} with 2 qubits.}
    \label{fig:qubatch}
\end{figure}

\subsection{QuBatch: Data Batch on Quantum Circuits}

\subsubsection{Fundamentals.}~\hspace{1pt}
Before introducing details, we use an example to illustrate the ability of quantum computing to process data batches in parallel in
Figure \ref{fig:qubatch}.
We have two vectors $D_1$ and $D_2$, each of which contains two features. The weight matrix is $U(\theta)$.
There are two computations to be performed: $U(\theta)\cdot D_1$ and $U(\theta)\cdot D_2$.

\begin{figure}[t]
    \centering
    \includegraphics[width=0.48\textwidth]{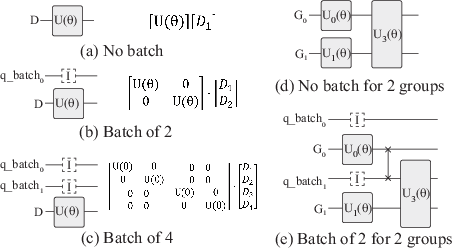}
    \caption{Illustration of integrating QuBatch to QuGeoVQC with different batch sizes and group sizes.}
    \label{fig:qubatchext}
\end{figure}

We construct a quantum circuit with three steps ($S_0$ to $S_2$) to perform these computations.
As shown in Figure \ref{fig:qubatch}(a), an encoder $U_e$ entangle these qubits from $S_0$ to $S_1$: $|\psi\rangle = U_e \cdot |0\rangle = [D_1,D_2]^T$.
The subcircuit from step $S_1$ to $S_2$ constructs the computation operator, where ``no gate'' is placed on qubit $q_0$, indicated by a virtual $I$, and the operator $U(\theta)$ is placed on qubit $q_1$.
As qubits are entangled at step $S_1$, we need to combine these two operators for computation by 
% to make the state transition from $S_1$ to $S_2$.
% The combination is based on the 
tensor product, shown in Figure \ref{fig:qubatch}(b).
Now, we have two $U(\theta)$ operators located on the diagonal of the matrix.
Kindly note that this indicates that we can duplicate the computation operator without any cost, providing the fundamentals to efficiently support SIMD operation.
Then, as shown in Figure \ref{fig:qubatch}(c), we can obtain the results of $U(\theta)\cdot D_1$ and $U(\theta)\cdot D_2$ on the output amplitudes.

\subsubsection{$QuBatch$ integration.}~\hspace{1pt}
To integrate $Qubatch$ to $QuGeoVQC$, we extend the approach in Figure \ref{fig:qubatch} from two aspects: (1) a larger number of batches and (2) the circuit with multiple groups.
For the first extension, $Qubatch$ can process batch size of $2^N$ by adding only $N$ qubits ($N\ge 0$), as shown in Figures \ref{fig:qubatchext}(a)-(c).
The second extension is to support multiple groups. 
This can be achieved by introducing a swap gate before communicating between two groups, as shown in Figures \ref{fig:qubatchext}(d)-(e).
We omit detailed analysis here to save space.
% for the sake of page limitations.

\subsubsection{Complexity analysis.}\label{sec:batch}~\hspace{1pt} Before detailed analysis, we first discuss the overhead introduced by $QuBatch$.
As multiple operators can be produced by using tensor products between one operator with an identity matrix, we do not need to pay overhead for the computation.
Kindly note that we need to pay overhead on the data encoding. 
First, the amplitude encoding will lead to a longer circuit.
Fortunately, as shown in \cite{li2023novel}, the circuit length grows linearly with the increase of qubits.
Second, because the sum of the squared magnitudes of the amplitudes of all possible states of a qubit must equal 1, if more data are encoded to a batch, we need to normalize the data, indicating that the data precision will be decreased.
Kindly note although the data precision will be decreased, the relevant relationship between data can be maintained.
In addition, we can control the group size and batch size to make a tradeoff among circuit depth, number of qubits, and data precision.

Now, we discuss the complexity.
Let $B$ be the number of batches implemented in \textit{QuBatch}, $G$ be the number of groups in the encoder, and $O(X)$ be the original time-space complexity (i.e., qubits times circuit depth).
By introducing $B$ bathes to the system, we will have overhead on both qubits number and depth: (1) the additional number of qubits will be $O(G\cdot \log B)$; (2) for each group, the circuit length increased along with the additional number of qubits; therefore, it has an overhead of $O(\log B)$.
As the encoding of different groups is conducted in parallel, we finally have the time-space complexity of $O(G\cdot \log^2 B\cdot X)$.
Compared to the implementation of $B$ batches independently, it will lead to a complexity of $O(B\cdot X)$.
If $B>>G$, \textit{QuBatch} can exponentially decrease the complexity.

% \clearpage

\section{Experiments}

\noindent This section reports experimental setups, followed by results.

\textbf{Dataset.}
Evaluations are performed on the FlatVelA dataset in OpenFWI \cite{deng2022openfwi}.
\textit{QuGeoData} is applied to adjust the data scale to fit the target quantum backends.
More specifically, the original data in OpenFWI has the dimension of $350,000 = 5\times 1000\times 70$~(\#source $\times$ time steps $\times$ \#receiver) for seismic data and $70\times 70$~(depth $\times$ width) for velocity maps. We set the constraint on the number of the qubits to be less than $16$, which fits most of today's superconducting-based or ion-trap-based quantum computers.
To this end, we will scale the dimension of seismic data to $256$, and the velocity map to $8\times 8$.
For comparison, we employ a standard nearest neighbor resampling algorithm to downsample both waveform and velocity map as the baseline, denoted as ``D-Sample''.
Two methods in \textit{QuGeoData} will be used for comparison.
We denote ``Q-D-FW'' to indicate the data (D) scaling using forward modeling (FW), which is proposed in Section \ref{sec:q-d-fw}, and denote ``Q-D-CNN'' to indicate the data (D) scaling using CNN-based data compression proposed in Section \ref{sec:q-d-cnn}.
% The backbone of ``Q-D-CNN'' is a 

\textbf{VQCs.}
We will compare two \textit{QuGeoVQC} designs with differences in the decoder.
The first one is designed to use the pixel-wise loss in Eq. \ref{eq:px}, where $8\times 8$ velocities will be decoded as the magnitude of 64 amplitudes from the quantum system.
We denote the design as ``Q-M-PX'', indicating the VQC model (M) based on the pixel-wise loss.
The second one is application-specific design, which leverages the knowledge that the subsurface has a flat layer-wise (LY) structure, denoted as ``Q-M-LY''.
It uses the loss function in Eq. \ref{eq:ly}, and the velocities will be decoded using Z-measurement of independent qubits.
Q-M-PX and Q-M-LY use the ansatz with 12 blocks, each of which is a `U3+CU3' block as proposed in \cite{wang2022quantumnas}.
To compare quantum learning and classical learning, we implemented two classical convolutional neural networks (CNN) with pixel-wise and layer-wise decoding, denoted as ``CNN-PX'' and ``CNN-LY''.
For a fair comparison, we use the same training and testing data and control the number of parameters of all models at the same level.

\textbf{Environment.} We carry out the concept-proof evaluation on quantum learning design based on the Torchquantum \cite{wang2022quantumnas} framework.
To enable the proposed encoder and \textit{QuBatch}, we added two components in TorchQuantum.
% to support the new group-based encoding and the process of with data batches.
For all VQC model training, we employ the Adam optimizer with 500 epochs where the initial learning rate is set to 0.1, followed by a cosine annealing schedule.
To support training, we split the FlatVelA dataset with 500 samples into a training set (size of 400) and a test set (size of 100).

The classical CNNs used in Q-D-CNN, CNN-PX, and CNN-LY are trained from scratch in Pytorch.
We used the same training setting (i.e., 500 epoch, Adam optimizer, and 0.1 initial learning rate, etc.).
For Q-D-CNN, we applied 500 other samples from the FlatVelA dataset to train the CNN.
Then, the trained CNN model is used to generate data for the test on quantum computing in Q-D-CNN.

\subsection{\textit{QuGeoData}: Physics Guidance is Needed}

\begin{figure}[t]
    \centering
    \includegraphics[width=0.48\textwidth]{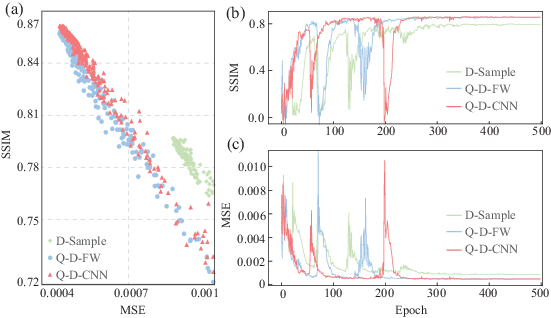}
    \caption{Performance of Q-M-PX VQC on dataset scaled by different approaches: (a) SSIM vs. MSE, the best solution is on the most left-top corner; (b) convergence comparison of SSIM in training; (c) comparison of MSE in training.}
    \label{fig:qudata}
\end{figure}

\noindent Figure \ref{fig:qudata} reports the comparison results of data scaling approaches using Q-M-PX VQC.
Each point in Figure \ref{fig:qudata}(a) is a VQC model obtained in training, where the x-axis and y-axis stand for Structural Similarity Image Metric (SSIM) and Mean Squared Error~(MSE), respectively.
Figures \ref{fig:qudata}(b)-(c) give the convergence of model training.

We have a couple of observations from these results.
First, we can clearly observe that the model trained based on the physics-guided data scaling (i.e., Q-D-FW) significantly outperforms D-Sample (i.e., green diamond) in both SSIM and MSE.
Second, Q-D-FW and Q-D-CNN have similar performance in two metrics.
Specifically, the SSIM of Q-D-CNN is 0.8619, which is even slightly higher than Q-D-FW with 0.8591, while the MSE of Q-FM is 0.000461, slightly outperforming Q-D-CNN with 0.000460.
The above results emphasize the importance of the incorporation of physics knowledge in data scaling.
Furthermore, it reflects that the learning-based approach can effectively scale data to keep physics features, which will be very useful in processing the raw seismic data.

\begin{figure}[t]
    \centering
    \includegraphics[width=0.48\textwidth]{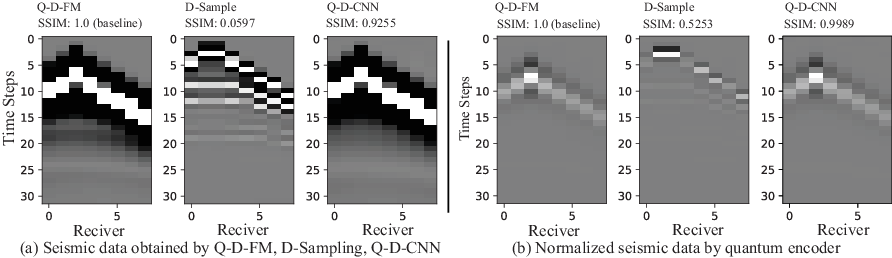}
    \caption{Visualization of seismic data by different methods: (a) scaled classical data; (b) normalized quantum data.}
    \label{fig:wavevis}
\end{figure}

Figure \ref{fig:wavevis} gives the visualization of seismic waveform data to better show the causes of performance gain.
First, in comparing Q-D-FM and D-Sample, we can see a larger wavelength in Q-D-FM.
The adjustment of the sampling rate and source wavelet causes this.
By down-scaling the time dimension in seismic waves, the sampling rate is decreased.
Consequently, we lower the source wavelet frequency from 15Hz to 8Hz to prevent the loss of information from a physics standpoint, resulting in an increased wavelength.
Conversely, D-Sample's approach of directly downsampling the waveform data leads to the inevitable loss of vital physical information, evident in the waveform measurements' incoherence, which will eventually degrade the resulting imaging accuracy. Secondly, the waveform data from Q-D-FM and Q-D-CNN, with an initial SSIM of 0.9255, see this value rise to 0.9989 following data normalization within quantum state constraints. These findings underscore the significance of physics-guided data scaling and the efficiency of CNN-based data compression methods.

\begin{figure}[t]
    \centering
    \includegraphics[width=0.48\textwidth]{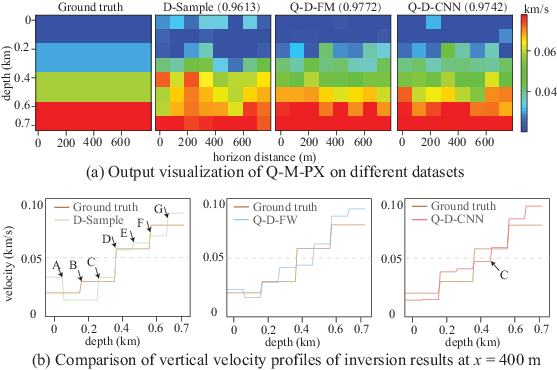}
    \caption{Visualization of the predicted velocity map: (a) outputs from different data scaling approaches; (b) vertical velocity profiling for physics information analysis.}
    \label{fig:exp1velocity}
\end{figure}

Finally, in Figure \ref{fig:exp1velocity}, we present the visualization of results.
% from Q-M-PX based on these data scaling approaches.
From Figure \ref{fig:exp1velocity}(a), we can again see the improvements obtained by Q-D-FM and Q-D-CNN over D-Sample.
More interesting results are shown in Figure \ref{fig:exp1velocity}(b), where we compare the vertical velocity profiles at horizon distance $x=400$.
The brown lines are the velocity changes along depth for the ground truth, where the y-value stands the velocity, and the layer boundary (say point B) indicates an interface between two layers.
From these figures, we can see that D-Sample leads to a larger difference in velocity values, and it cannot predict the interface.
Specifically, for all seven inflection points, there are five wrong predictions (points A, B, C, E, and G) and only two correct predictions (points D and F).
On the other hand, there are three correct interface predictions for both Q-D-FW and Q-D-CNN.

Clearly, the aforementioned results indicate that creating a dataset in accordance with physical principles can enhance performance. However, to fully leverage quantum computing for FWI problems, further improvements in performance are necessary.

\subsection{\textit{QuGeoVQC} Further Boost Performance}

\begin{figure}[t]
    \centering
    \includegraphics[width=0.4\textwidth]{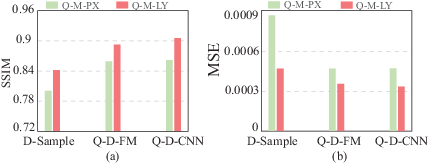}
    \caption{Comparison results of Q-M-PX and Q-M-LY.}
    \label{fig:qumodel}
\end{figure}

\noindent Figure \ref{fig:qumodel} reports the comparison results on two different VQC designs: Q-M-PX and Q-M-LY, using both SSIM  and MSE metrics.
As shown in the figures, for all three data scaling methods, results obtained by Q-M-LY significantly outperform those by Q-M-PX.
Specifically, for SSIM using data from Q-D-CNN data, the SSIM improves from 0.862 to 0.905.
These figures are from 0.859 to 0.892 for Q-D-FW, and 0.800 to 0.842 for D-Sample.
Overall, Q-M-LY achieves a 4.5\% improvement over Q-M-PX.
For MSE, the average improvement reaches 33.23\%.
These results show the effectiveness of layer-wise VQC design.

We have one more interesting observation from the results.
Considering that the Q-M-PX using D-Sample is the straightforward implementation of FWI on quantum computing, the proposed \textit{QuGeo} design can improve the SSIM from 0.800 to 0.905 and reduce MSE from 0.000855 to 0.000328, achieving 11.6\% and 61.69\% improvements on SSIM and MSE, respectively.
This result shows the huge optimization room for implementing FWI on quantum computing, and the effectiveness of our proposed methods.

\begin{figure}[t]
    \centering
    \includegraphics[width=0.48\textwidth]{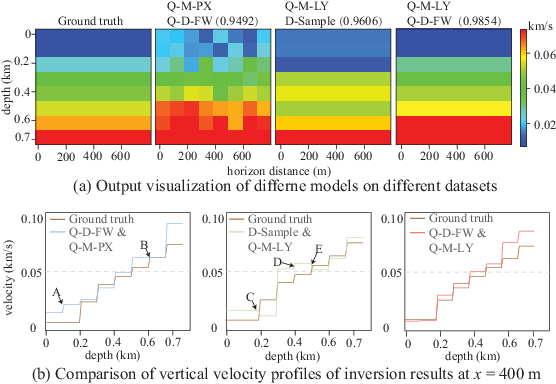}
    \caption{Visualization of the predicted velocity map for Q-M-LY: (a) ground truth and prediction results; (b) vertical velocity profiling for physics information analysis.}
    \label{fig:exp2vis}
\end{figure}

Figure \ref{fig:exp2vis} presents the visualization results for Q-M-PX and Q-M-LY using Q-D-FW, as well as Q-M-LY using D-Sample.
In the vertical velocity analysis depicted in Figure \ref{fig:exp2vis}(b), we observe that Q-M-PX inaccurately predicts two interfaces at Points A and B.
Although Q-M-LY using D-Sample accurately predicts all interfaces, it incorrectly interprets the relative positioning between two layers at three interfaces, as highlighted at Points C, D, and E.
In contrast, Q-M-LY using Q-D-FW, successfully predicts all interfaces while maintaining the correct relative relationships between the layers.
These findings further emphasize the effectiveness of layer-wise VQC design and highlight the promising potential of quantum learning in addressing the challenges of FWI.

\subsection{\textit{QuBatch} Achieves Competitive Results}

\begin{table}[t]
    \centering
    \begin{tabular}{|c|c|c|c|c|c|}
    \hline
        Model & Dataset & Batch & Extra Qubits & SSIM & vs. BL  \\ \hline
        \multirow{3}{*}{Q-M-LY} & \multirow{3}{*}{Q-D-FW} & 0 & 0 & 0.8926 & BL  \\ 
        ~ & ~ & 2 & 1 & 0.8864 & 0.69\%  \\
        ~ & ~ & 4 & 2 & 0.8678 & 2.77\%  \\ \hline
    \end{tabular}
    \caption{Evaluation of \textit{QuBatch} with different batch sizes on Q-D-FW dataset using Q-M-LY VQC.}
\label{tab:bigtable}
\end{table}

\noindent Next, Table \ref{tab:bigtable} shows the proposed \textit{QuBatch} can be successfully used in the model training, where we can perform $2^N$ batches in parallel with only $N$ more qubits.
In addition, we observe the SSIM has slight degradation compared with the results obtained without using data batch.
The root cause might be the decrease in data precision caused by the normalization, which is required by the constraints of amplitudes.
% , as analyzed . 
As discussed in Section \ref{sec:batch}, such effects can be mitigated by making tradeoffs among data precision and qubits, which will be the future work.

% We do not perform tradeoff optimization in this work, which will be the future work.

\subsection{\textit{QuGeo} Outperforms Classical ML}
% Table generated by Excel2LaTeX from sheet 'Sheet8'
\begin{table}[t]
  \centering
  \tabcolsep 1 pt
  \renewcommand{\arraystretch}{1.25}
  \small
  \caption{Comparison between quantum and classical learning}
    \begin{tabular}{|c|c|cccc|cccc|}
    \hline
    \multirow{2}{*}{Model} & \multirow{2}{*}{Par.} & \multicolumn{4}{c|}{Q-D-FW}    & \multicolumn{4}{c|}{\textbf{Q-D-CNN}} \\
\cline{3-10}          &       & SSIM  & vs. & MSE  & vs. & SSIM  & vs. & MSE   & vs.  \\
    \hline
    CNN-PX & 634   & 0.870  & BL    & 4.34E-04 & BL    & 0.87  & BL    & 4.38E-04 & BL\\
    CNN-LY & 616   & 0.871  & 0.04\% & 4.36E-04 & -0.43\% & 0.87  & 0.00\% & 4.36E-04 & 0.38\% \\
    Q-M-PX & 576   & 0.859  & -1.28\% & 4.61E-04 & -6.10\% & 0.86  & -0.98\% & 4.62E-04 & -5.45\% \\
    \textbf{Q-M-LY} & \textbf{576} & 0.893  & \textbf{2.50\%} & 3.48E-04 & \textbf{19.84\%} & 0.91  & \textbf{3.87\%} & 3.28E-04 & \textbf{25.17\%} \\
    \hline
    \end{tabular}%
  \label{tab:qvc}%
\end{table}%

% Last but not least, 
\noindent Table \ref{tab:qvc} reports the comparison of our proposed quantum learning for FWI with classical learning.
We constrain the number of parameters to the same level.
As shown in Table \ref{tab:qvc}, CNN-PX and CNN-LY have 58 and 40 more parameters.
Whereas, Q-M-LY outperforms both classical results on a dataset obtained by both Q-D-FW and Q-D-CNN.
Specifically, using CNN-PX as a baseline, Q-M-LY achieves 19.84\% and 25.17\% MSE improvements on two datasets, respectively.
The potential reason that quantum learning can outperform classical learning at the same scale is because of the complicated entanglement in quantum computing, which has the potential to extract high-correlation features among data.
Achieving better performance even at the small scale is meaningful since real-world applications commonly have hard real-time requirements, in particular, for monitoring tasks in geoscience. 

% \clearpage
\section{Conclusion}
\noindent This pilot study introduces \textit{QuGeo}, a framework designed to facilitate full-waveform inversion (FWI) tasks from geoscience using near-term quantum computers.
We employ machine learning to tackle two fundamental challenges in applying quantum computing to FWI: (1) identifying the most suitable data for quantum computing, and (2) understanding the unique advantages offered by quantum computing in this context.
With \textit{QuGeo}, we have developed a physics-guide quantum data scaling tool. Additionally, we propose a novel approach for data batch execution on variational quantum circuits, which could potentially enable efficient training with large data batches.
Notably, our quantum learning model outperforms its classical counterpart in performance, even with an equal number of parameters.
Furthermore, \textit{QuGeo} achieves an impressive 61.69\% improvement in MSE compared to conventional quantum learning approaches. We believe that this research paves the wave for practical applications of quantum computing in geoscience, potentially revolutionizing the field.

\vspace{-2pt}
\bibliographystyle{unsrt2authabbrvpp}
\bibliography{paper}

\end{document}